\documentclass[a4,12pt]{article}
\usepackage[2cell,arrow,curve,matrix,web]{xy}
\UseHalfTwocells \UseTwocells
\usepackage{graphicx}
\begin{document}

\begin{titlepage}




\vspace{.5cm}

\begin{center}

{\Large \bf Neural Graphs \&\ Category of Memory States }

\vspace{1cm}

{\large I. Mitra{\footnote{indranil@theory.saha.ernet.in}}$^a$}
\vspace{5mm}

{\em $^a$Department of Physics, New Alipore College\\
L Block, Kolkata 700 053, India}\\

\vspace{.5cm}
\end{center}

\vspace{.5cm}

\centerline{{\bf{Abstract}}}

\vspace{.5cm}

\begin{small}
The brain as an astonishingly remarkable device has been studied
from various angles. It is now well known that neurons are the
seat of all activities of the brain function. The dynamical
properties pertaining to a single neuron and a collection of
neurons may be widely different owing to the clustering
properties of a group of neurons. As it can be clearly understood
theory of complex physical systems has been more and more
employed to study the behaviour of neurons and neuronal circuits.
We here mainly discuss neural correlates of memory and cognitive
functions utilizing graph theory and ideas from geometry. It has
been suggested that stochastic processes being at the helm of
affairs in the neuronal level there may exist surfaces to some
extent like a hologram for the existence of memory functions.It
is also instructive to mention that Amari's developments
 \cite{amari} as regards information geometry has acted as an
important inspiration. Unlike some previous analysis
categorization of memory from neural perspectives have been
reconsidered at the neuronal level. In essence the main point of
discussion here has been to give an alternative model of memory
where stochastic geometry and algebraic surfaces is an important
ingredient.

\end{small}

\end{titlepage}
\newpage

\section{Introduction}
Our current understanding of neuroscience is mainly concerned with
the studies of brain as a complex object consisting of a systems
of neurons whose proper dynamics is still unknown though several
models have been proposed. A brain event may be characterized by
an activation which moves through the brain as neurons act
sequentially in some physical process.Among these models the
Hopfield \cite{hop} and its generalizations have been extensively
studied and an important result has been obtained like the famous
relation $p/N \approx 0.14$ where N is the number of neurons and
p are the stored patterns \cite{amit}. The general rule of thumb
assumed in these models is that the changes in the synaptic
strengths are proportional to the correlation between firing of
pre and postsynaptic neurons. Since the volume density of
synapses in gray matter is brain size independent\cite{sch}, i.e.
$NM/V_{g}= const$, then we obtain that $M\sim V_{g}/N$, and as a
consequence $p\sim V_{g}/N^{2}$. The total number of neurons $N$
is proportional to the total cortical surface area $W$
\cite{rock}. The latter scales with the brain volume as: $W\sim
V_{g}^{0.9}$ for large convoluted brains \cite{hof}. This leads to
the following scaling between the average connectivity $p$ and
brain size for convoluted brains.
\begin{equation}
p\sim V_{g}^{-0.8}.
\end{equation}\\
Studies have also indicated \cite{gross} that the human brain is a
large system, with no more than a hundred specialized modules with
different functions. At the fundamental level, the cerebral cortex
consists of about $10^{10}$ neurons that comprise a highly
interconnected network. Each cell receives continuously a few
thousands of excitatory inputs from other neurons. One of the
basic things which is not known is how the cortex, being a mainly
excitatory network, prevents the expected explosive propagation
of activity and still transmits information across areas. To be
more concrete, brain activity happens in bursts, in which pauses,
silence or inactivity suddenly and unpredictably are followed by
transient activity. The probability $Q$ that the area X connects
with the area Y is exactly complementary to the probability that
none of the modules in X connects with the area Y, i.e.
$(1-\kappa)^{W_{0}/\xi^{2}}$, where $W_{0}/\xi^{2}$ is the number
of modules in every area. Thus,  $Q$ is given by
\begin{equation}
Q\approx 1 - \exp\left(- \frac{a qL_{0}^{2}}{\xi^{2}K^{2}}\right).
\end{equation}\\
where  $L_{0}$ is the average length of axons in white matter,
$K$ is the number of areas in the cortex, $\kappa$ is the
probability of connection between a given module in one area to
another area, $\xi$ is the linear size of a module in the
cortex,  $a$ is a dimensionless parameter characterizing cortical
geometry and a pattern of axonal organization in white matter,
$q$ is the probability of sending at least one macroscopic axonal
bundle to white matter by a module,  $W_{0}$ is the surface area
of one cortical area. It is also possible to find an expected
number of modules in one area that connect with another area.
Assuming that modules are statistically independent, i.e. the
probability of sending axonal bundles for a given module does not
depend on other modules, the distribution of the number of
modules in A reaching B is represented  by a binomial
distribution. Thus the average number of modules in A connecting
with B is given by the product of the probability that a module
in A connects with area B  and the number of modules in A
($W_{0}/\xi^{2}$), i.e. $aqL_{0}^{2}/\xi^{2}K^{2}$. From a
dynamical perspective brain dynamics is not different from other
natural processes. Nature is clearly non homogeneous and
intermittent, the analysis of any natural object reveals an ever
surprising amount of details, there is no single relevant scale at
which Nature becomes homogeneous. Complexity is this lack of
uniformity associated with the scale-free spatiotemporal feature.
It is now widely recognized that, under a variety of conditions,
non linear systems with many degrees of freedom tend to evolve
towards complexity and criticality \cite{bak}.  It is the
interaction of many nonlinear degrees of freedom which produces
emergent complex dynamics. Brain activity is eminently
spatio-temporal, as such the monitoring of the complicated
cortical patterns have greatly benefited from techniques
developed in the context of functional magnetic resonance imaging
(fMRI). However, the numerical analysis of such spatiotemporal
patterns is less developed, lacking mathematical tools and
approaches specifically tailored to grasp the complexity of brain
cortical activity. One possibility is to get insight from recent
work showing that disparate systems can be described as complex
networks, that is assemblies of nodes and links with nontrivial
topological properties \cite{top,top1}. The brain creates and
reshapes continuously complex functional networks of correlated
dynamics responding to the traffic between regions, during
behavior or even at rest. Some recently studied networks, using
functional magnetic resonance imaging in humans \cite{band} has
been done. The data is analyzed in the context of the current
understanding of complex networks. Some statistical properties of
these networks, are, path length and clustering. The path length
($L$) between two brain sites is the minimum number of links
necessary to connect both of them. Clustering ($C$) is the
fraction of connections between the topological neighbors of the
sites with respect to the maximum possible. The average
clustering of a network is given by $C = 1/N\sum_i C_i$, where
$N$ is the number of sites. The scale-free features reflects
underlying long range correlations, i.e., brain activity on a
given area can be correlated with far away and apparently
unrelated regions, something already documented with other
technology \cite{link}. One gains a lot of insight into the
spontaneous magnetic ordering below a critical temperature if one
studies the Ising model, which replaces the rather complex
ferromagnetic atom by a simple binary unit interacting with its
neighbors. With this analogy it is definitely useful to
investigate simple units, which model a few essential mechanisms
of neurons and synapses, and to  study the cooperative behaviour
of such interacting units. It is not obvious at all, whether such
a system can store an infinite number of patterns with one set of
synapses, learn from examples and generalize. It will be
important here to mention that stochastic behavior is crucially
important in this analysis and the onset of bifurcation and chaos
will be an important determining factor based on the studies of
Lyapunov exponents. The neural network of the human brain
responds as a unified whole memory bank to a multitude of input
signals from the environment and functions with a high degree of
robustness and stability. The three aspects of neural networks
memory bank are, storage, real-time update and retrieval. The
memory is believed to be embedded in the strength of the numerous
connections or synapses in the network. Sensory inputs
(electrical) produce particular patterns of activity in groups of
neurons which then trigger optimal response to the input signal.
The cooperative response of millions of neurons to a multitude of
input signals has been compared to a very efficient parallel
processing computer with neurons and their synaptic connections
as fundamental units of information processing, like switches
within computers. However, recent studies \cite{hamm,ras} show
that neurons and synapses are extremely complex and resemble
entire computers, rather than switches. The interiors of neurons
are now known to contain highly ordered parallel networks of
filamentous protein polymers collectively termed the
cytoskeleton. Information storage, update and appropriate
retrieval are controlled at the molecular level by the neuronal
cytoskeleton which serves as the internal communication network
within neuron. Organization of information at the molecular level
in the cytoskeletal network contributes to the overall response
of each neuron and the collective activity pattern of neurons
then governs the response to the environmental stimuli.

The general awareness or evolution of cognitive functions of the
individual may also be governed by the overall background activity
pattern of the neurons and their cytoskeletal networks. Coherent
signal flow patterns in neural networks may form the basis for
general consciousness and response to stimuli (external or
internal). Inputs signals trigger spontaneous appropriate
coherent pattern formation in the activity of the neurons with
implicit spatial correlations in the activity pattern. The time
variation of electrical activity of the brain as recorded by the
Electro Encephalogram (EEG) exhibits fluctuations on all scales
of time, i.e. a broadband spectrum of periodicities (frequencies)
contribute to the observed fluctuations \cite{eeg}. Power spectral
analysis which is used to resolve the component frequencies (f)
and their intensities, shows that the intensity (power) of the
component frequencies follow the inverse power law form $1/f^{B}$
where B is the exponent. Inverse power law form for power spectra
of temporal fluctuations imply long-range temporal correlations.
The signatures of short - term fluctuations are carried as
internal structures of long - term fluctuations.Neural network
activity patterns therefore exhibit long - range spatial and
temporal correlations. Such non-local connections in space and
time are ubiquitous to time evolution of spatially extended
dynamical system in nature and is recently identified as
signature of self-organized criticality \cite{crit}. Extended
dynamical systems in nature have selfsimilar fractal geometry.
Selfsimilarity implies that submits of a system resemble the
whole in shape. The fractal dimension D is given by $\frac{d ln M
}{d ln R}$ where M is the mass contained within a distance R from
a point within the extended object. A constant value for D
implies uniform stretching on logarithmic scale for length scale
range R.
  The association of fractal structures
with chaotic dynamics has been identified in all dynamical
systems in nature. The computed trajectory of time evolution
exhibits fractal geometry. The branching interconnecting networks
of neurons and intra-neuronal cytoskeleton networks are fractal
structures which generate electrical signal pattern with
self-similar fluctuations on all scales of time characterised by
1/fB power law behavior for the power spectrum. Such inverse
power law form for spectra of temporal fluctuations implies
long-range temporal correlations, i. e., long term memory of
short term fluctuations or events. Fractal architecture of neural
networks supports and coordinates information (fluctuations) flow
on all time and space time scales in a state of dynamic
equilibrium, now identified as self-organized criticality, is
ubiquitous to natural phenomena and is independent of the exact
details of the dynamical processes governing the space-time
evolution. The physics of self-organized criticality or
deterministic chaos is not yet known. Among some other
interesting works \cite{pib} pioneering work, showed that many
functional activities of the brain involve extended assembly of
neurons. On this basis, some concepts of Quantum Optics, such as
holography, in brain modeling has been developed. Information is
indeed observed to be spatially uniform in much the way that the
information density is uniform in a hologram \cite{fre1}. While
the activity of the single neuron is experimentally observed in
form of discrete and stochastic pulse trains and point processes,
the ``macroscopic'' activity of large assembly of neurons appears
to be spatially coherent and highly structured in phase and
amplitude . The quantum model of brain proposed \cite{ur} is
firmly founded on such an experimental evidence. The model is in
fact primarily aimed to the description of non-locality of brain
functions, especially of memory storing and recalling. The
mathematical formalism in which the model is formulated is the
one of Quantum Field Theory (QFT) of many body systems. The main
ingredient of the model is thus the mechanism of spontaneous
breakdown of symmetry by which long range correlations (the
Nambu-Goldstone (NG) boson modes) are dynamically generated in
many body physics. In the model the "dynamical variables" are
identified \cite{ya} with those of the electrical dipole
vibrational field of the water molecules and of other
biomolecules present in the brain structures, and with the ones
of the associated NG modes, named the dipole wave quanta. The
model, further developed exhibits interesting features related
with the role of microtubules in the brain activity \cite{pen}
and its extension to dissipative dynamics allows a huge memory
capacity. The dissipative quantum model of brain has been
investigated \cite{pv} also in relation with the modeling of
neural networks exhibiting long range correlations among the net
units. One motivation for such a study is of course the great
interest in neural network modeling, in computational
neuroscience and in quantum computational strategies based on
quantum evolution \cite{qc}. Among some other important
contributions utilization of functional geometry and an approach
of statistical manifold in the brain domain(CNS) \cite{roy} has
drawn considerable attention in recent years. The proposal is
based on the emergence of cognition on the metric and statistical
properties of the manifold.

Memory is believed to be a universal feature of the nervous
system  and exciting results improving our understanding of
molecular as well as organizational mechanisms underlying memory
have been obtained in recent decades \cite{mem}. On the
organizational level significant work has been devoted to the
study of ``brain maps'' underlying the ability to recognize
patterns or features from a given sensory input \cite{map}. Many
intriguing suggestions have been given as to how a memory emerges
that is able to extract and recall features from a spatial
pattern of neural activity. Time is important in many cognitive
tasks but the crucial point is how to represent time, and methods
often involve time delays in one form or another. How does a
structured memory emerge that can cope with temporal sequences of
activity? For example, the information we receive through a
temporal sequence of input must at least to some extent be
memorized spatially in the neuronal activity pattern.
Illustration of these features has been considered in \cite{comp}
on the fundamental assumptions (a) {\em Competition} between
neural units where excited neural units have an inhibiting effect
on other units. (b) {\em Hebbian Plasticity} is an abstract
formulation of long term potentiation depending on pre- and
postsynaptic activity: If activity of unit A is followed by
activity of unit B the connection from A to B is strengthened
\cite{hebb}. (c) {\em Recurrent} connectivity opens up the
possibility for ongoing information processing in the network by
internal feedback. Recently a model for quantum channels with
memory has been proposed that can consistently define quantum
channels with Markovian correlated noise \cite{markov}.  The model
also extends to describe channels that act on transmitted states
in such a way that there is no requirement for interaction with an
environment within the model.

 So it should be clear that to study
the cognitive functions and memory aspects in brain surprisingly
geometry may be an major ingredient. As far quantum mechanics is
concerned it is not at present clear that at what scale it may be
operative \cite{roy1}. The present work is mainly an approach to
the problem of modeling memory and cognitive states from a
kinematical or operative point of view, in a sense that we at
present being unaware of the exact dynamics of neurons it would
be helpful to see and speculate what actually happens. The model
we propose here to some extent has a overlap between the
holographic and neuromanifold pictures. But the major difference
is that they have an algebraic flavour which we will see shortly.
The proposal  \cite{catg} implies that there may be an inherent
categorization of the inputs, though it is possibly difficult at
this stage to comment on the mechanism. We are of the opinion
that there may be different viewpoints and perspectives in
studying neurobiology but as far understanding the brain and
trying to construct a dynamical theory of the neural mechanism
geometry may inevitably play a very important role.

The short paper is organized as follows. In the next section we
develop the basic model which we intend to explore. Thereafter we
speculate some important connections with the real systems and
try to devise a geometric picture.

\section{ Neuron Sites and Signal Processing Graphs}

Signal processing is inherently stochastic \cite{mon}. It would
not be an overestimation to state that neural activities do employ
signal processing. But the most difficult to answer would be how
does it perform?. We show below a diagram to show how impulse can be propagated through the axon and the dendrites.
\begin{center}
\begin{figure}[htbp]
\centering
  \includegraphics{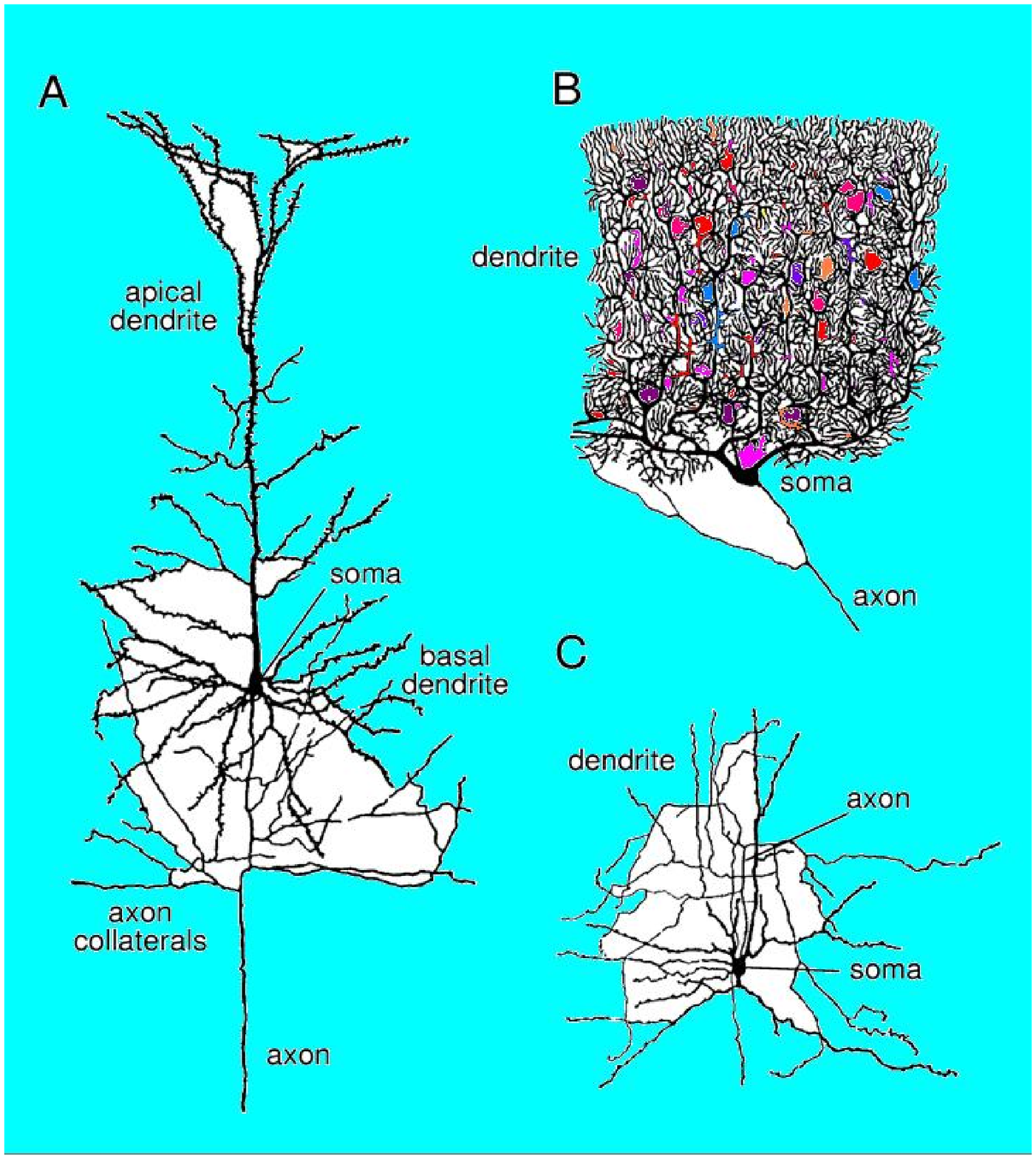}
    \label{fig:Picture3}
    \end{figure}
\end{center}
Loosely speaking, the diagram (B) above resembles a tree graph.
In general information proccesing by the neurons can be modeled
as a propagation of signal as a transmission line along the nerve
cells by a diffusion like process for the voltage transmission $$
b_{k}\frac{\partial V}{\partial t} = {\mathcal I}_k +
\frac{1}{d_{k}}\frac{\partial}{\partial x}(s \frac{\partial
V}{\partial x})$$ for some choice of the parameters and current.
So essentially information processing here arises out of the
consideration of the neurons being elements of a circuit.
For example the Hodgekin-Huxley model can be reagarded as realising a neuron as an RC circuit on the ionic channels
\cite{hodge}.
\begin{figure}
    \centering
        \includegraphics{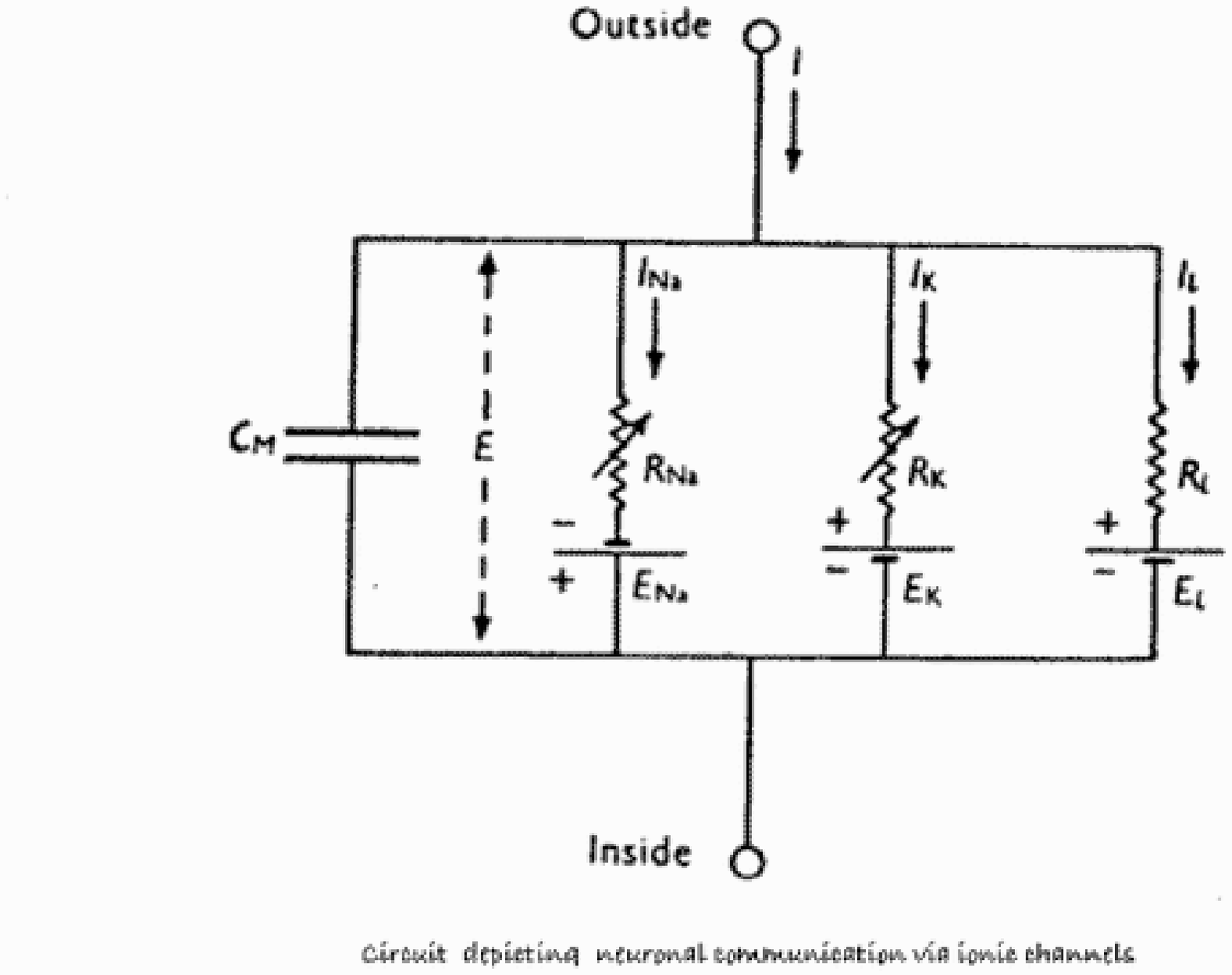}
    \label{fig:pic33}
\end{figure}
But the
essential point of concern here is whether the processes are
stochastic or not. We will delve into these questions in detail
soon but at present we would like to mention that there exists
some important results which predict neuronal behaviour with some
network assumptions with $$ V(t) = \int^{0}_{t} \mathcal{K}(t -
\zeta) d_{k}{\mathcal I}_k (\zeta) d\zeta$$ Where
$\mathcal{K}(\psi)$ is the kernel and will depend on the chosen
model. In general for a spatially structured spike-response model(SRM)
neurons the analysis is simpler. This analysis is also very important as our analysis has some structural similarities
with it.
If we have a large number of SRM neurons arranged on a two-dimensional grid. The synaptic coupling strength $A_{ij}$ is
 a function of their distance in a functional space of the input impulses $\lambda_{i}^{a}$. The response of a neuron to
 the firing of one of its presynaptic neurons is described by a response function $\epsilon$ and, the potential is given
 by
  a kernel $\mathcal{K}$. The membrane potential of a neuron located at $x_{i}$ is given by
$$V(\lambda_{i}^{a}, t) = \int_{0}^{\infty} d\psi {\mathcal K}(\psi)
 S(\lambda_{i}^{a}, t - \psi) +  \sum _{j} A_{ij}\int_{0}^{\infty} d\psi  \epsilon
  S(\lambda_{i}^{a},  t - \psi)$$
with $S(\lambda_{i}^{a},  t)$ is  the spike train of the neuron.
\begin{figure}[htbp]
    \centering
        \includegraphics{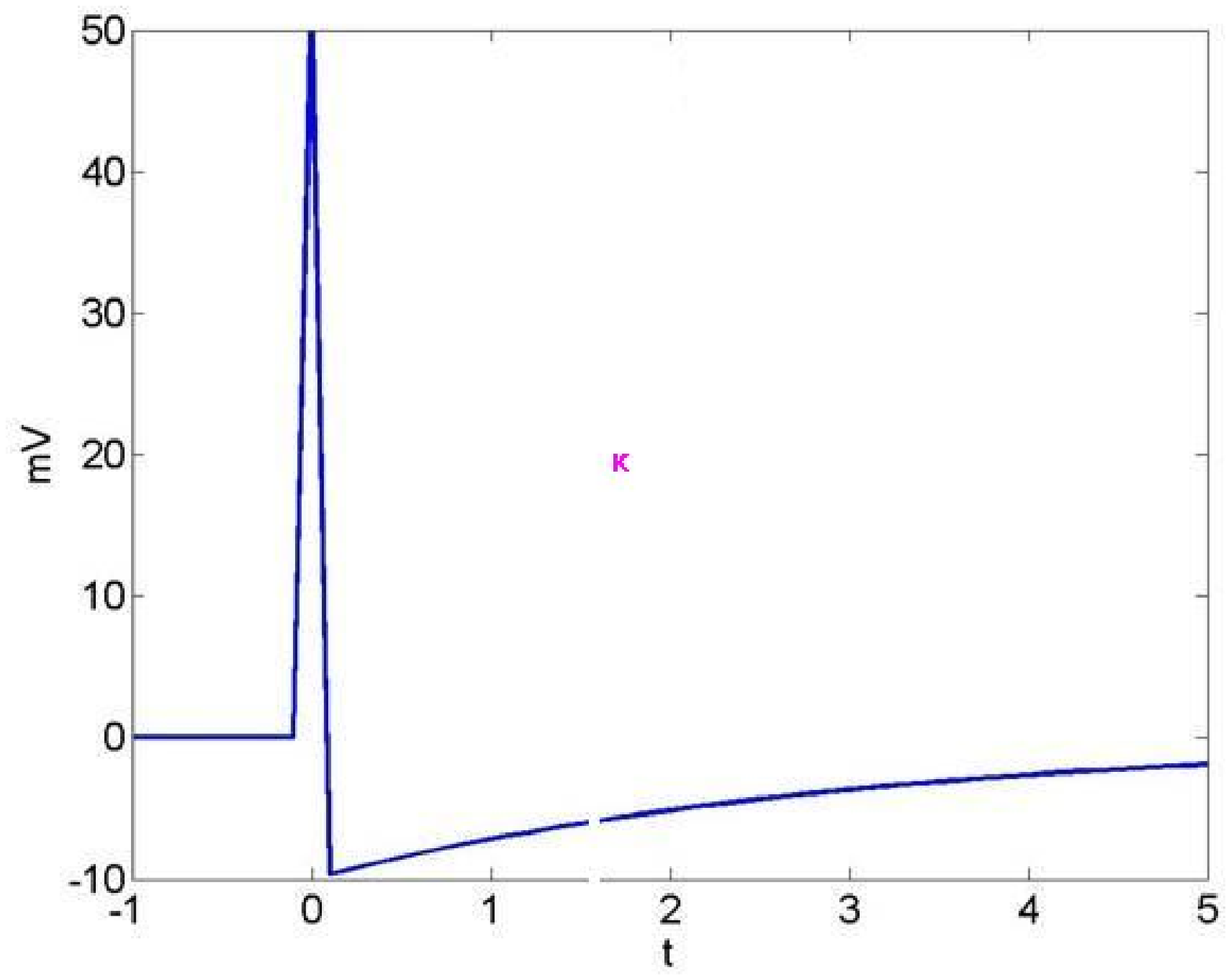}
        \label{fig:Picture11}
\end{figure}
In communication channels every input signal along
with a weight factor propagates along with a noise tolerance.
There are some ANN models which employ such mechanism. It is
important to realize the generation and propagation of a nervous
impulse. When the threshold of stimulus is reached the system is
switched over to a new state. So as a whole though it is not
clear but it is quite tempting to think that the neuronal
processes are essentially monitored by collection of neurons. In
our proposal we consider a collection of afferent, intermediate
and efferent neurons.
It should be mentioned that we at present
are ignoring the signal processing aspects of the neurons as we
will see shortly that we being mainly interested in memory,
though not completely justified those things may be ignored for
the present.At this stage the primitive assumptions of the model
should be particularly mentioned. It is assumed in our model that
the intermediate neurons are large in number in comparison of the
other two. It should be mentioned at this point that they are
functional attributes and may have some implications on the model
after experimental findings. Next we assume that the neurons are
arranged and interconnected as if on a lattice, though this may be
a debatable issue physiologically, but again we state here that
this should not be thought as a physical picture of the
descriptions of the neurons but rather a functional point of
view. The question that may be asked at this stage whether it is
justified to assume a structure may it be functional from the
very onset. In this context we wish to state that the mappings of
the brain function as has been recently confirmed by some MEG
experiments \cite{meg} to the cortical area is topological. But we
are yet to underpin the nature and form of the mapping.

\subsection{Relationships of Input and Output Patterns and Categorization}
Let us consider a input pattern {\footnote{though we neglect here
some crucial questions to be asked here as how can we say an
initial input to be a pattern before the information is
processed}} $\lambda^{\varepsilon}_{i}$ and the output pattern
$\psi^{\varepsilon}_{j}$ The meaningful question which are
relevant to be asked for this is the probability
$P(\psi_{j},\lambda_{i})$ with its usual connotations. As is well
known from statistical mechanics the average information content
for this given configuration is evaluated to be
\begin{eqnarray}
\Big< I(\psi, \lambda)\Big> = \lim_{n\rightarrow
0}\frac{1}{n}\Big< \int d\psi d\lambda P( \psi, \lambda)
\big([\frac{P( \psi, \lambda)}{P(\psi)}]^{n} -
[P(\lambda)]^{n}\big)\Big>
\end{eqnarray}
 The above integrals for
large neuronal systems is obtained by saddle point evaluation and
they may imply a correlation with the signals.

\begin{figure}[htbp]
    \centering
        \includegraphics{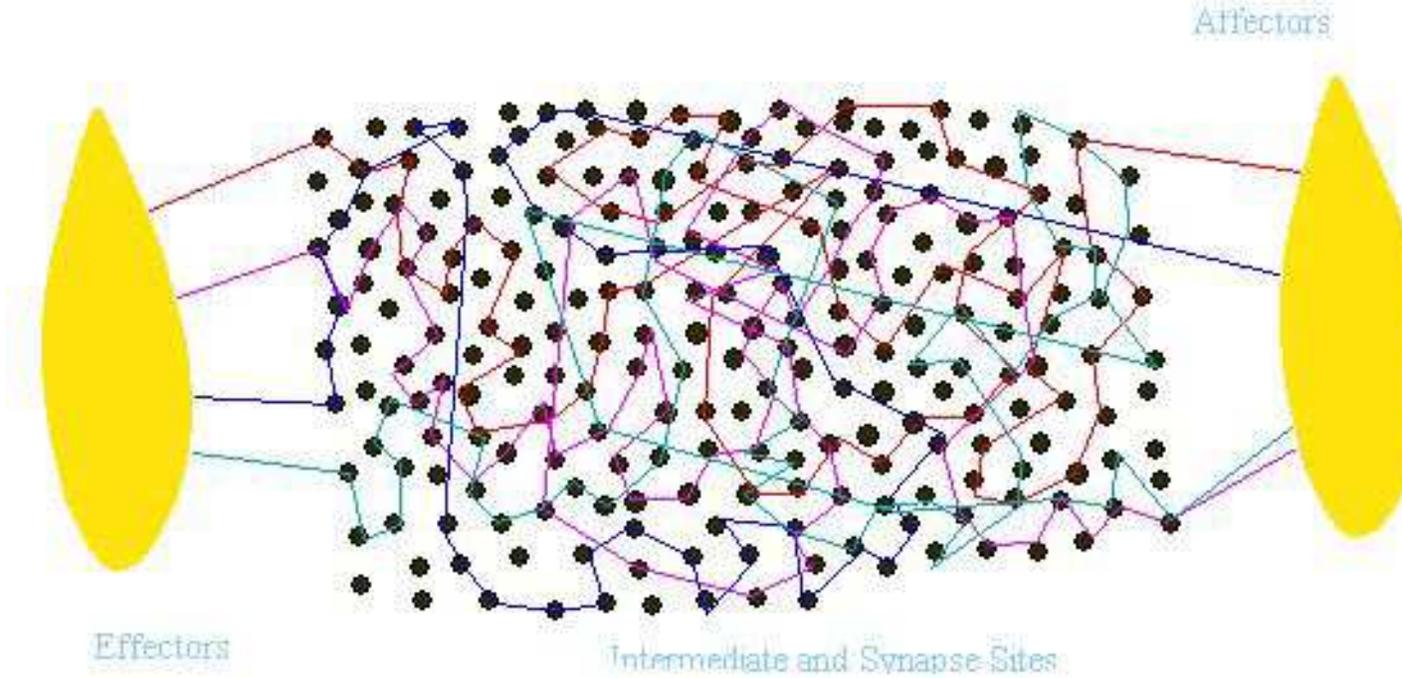}
    \caption{Neuronal Graphs}
    \label{ng}
\end{figure}

Now as we have previously assumed about or proposed model
regarding neurons as input output systems on lattice sites, the
essential ingredient of our assumption is that cognitive aspects
are a result of dynamics and clustering of neurons in various
alignments and arrangements. To depict the picture of the model
in detail the lattice sites are occupied by synapse with weights
$ A_{ij}$ determined by the position of the lattice sites. We
think of the neurons as functionally connected together by the
synapses, situated at different sites. Now any sensory experience
as an input $(\lambda^{a}_{k})$, from a dynamical point of view
gives rise to a spike $(e_{ijkl})$ (connecting the ij'th site to
the kl) which interconnects a group of neurons from the afferent
to the intermediate and again with a variety of possibilities
connects the efferent ones which gives rise to outputs
$(\psi^{a}_{k})$. In general the probability distribution of a
graph is associated with an input and output state. So in general
for a path from the effectors to the affectors via the
intermediate layer of neurons will in general be composed of
combination of these paths. So rigorously speaking the
probability density $P(\psi_{j},\lambda_{i})$ associated should
be taken between all the $n$ (say) input states. If we label the sites with
indices then the actual probability density for a effector to
affector takes the form of
\begin{eqnarray}
\mathcal P &=& \left(
\begin{array}[c]{clrrr}
  p_{11} & p_{12} & \cdots & \cdots & p_{1n} \\
  p_{21} & \cdots & \cdots & \cdots & \cdots \\
  p_{31} & \cdots & \cdots & \cdots & \cdots \\
  p_{m1} & \cdots & \cdots & \cdots & p_{mn} \\
\end{array}
\right )
\end{eqnarray}
Here we signify $p_{ij}$ as the probability density associated with a chain of the graph,
connecting the
$ij$ site. In our arguments below though we will mainly be concerned with the
probability density function $\mathcal P$ disregarding the chain probability densities.
It should be clarified at this stage that as wee will see in our
development that the neuronal processes do behave as a Markov
chain, and the states are given by an activation map $\varphi :
\lambda^{a}_{k} \rightarrow x$. But effectively here we in this
paper have made a simplifying choice on the function and made it
an identity. {\footnote{Various models of brain wave may be
analogous to this idea, it may be also thought the waves to
behave as solitons \cite{wav}.}}So in this model essentially any
input as wee see can be thought of as a possible groups of
neurons attached with proper weights of synapses at the sites,
and the outputs do form sensory experiences mainly memory from
our perspectives. Here it should be mentioned that we depict the
model here mainly which gives rise to cognitive states such as
memory, and ignore some other important aspects such as motor
outputs corresponding to the inputs. So importantly here we
assume that any experience is basically a network of neurons on
the functional lattice space and they can be thought as graphs.We
give in Fig (\ref{ng}) a simple diagram of a model graph of
neurons for a very low number of sites. It should be noted that
there are always possible to have different graph for the same
input and outputs, thereby giving rise to the notion of
probability distribution for the a specific graphs for a given
input-output. In the figure however we have shown some graphs
with different input and outputs.
The essential point here is that we consider the neural circuits
as digraphs ${\mathcal G}_{i,j}$ with n vertices and m edges. We
will later try to construct adjacency matrices corresponding to
these graphs to see how they can give nontrivial conclusions
regarding a network.
 We assume here that the graphs may not be unique in a
sense that corresponding to a group of similar set of experiences
we may charecterize the same graph. This can be labelled as one
of the limitations of the model, but it is emphasized here that by
introducing more parameters in the function space this problem
may be resolved. For example we may assume that the neurons have
some extra attributes attached to them. So the functional neuron
space is mainly composed of graphs which give rise to memory and
related cognitive phenomena corresponding to a set of
experiences. The graphs can also be thought of arising from
thinking the cognitive and memory states to be objects
$\mathcal{M}$ which consists of $(\lambda^{a}_{k}, \psi^{a}_{k})$
and the two maps which consists of $ A_{ij}$ and $e_{ijkl}$. So this
can be represented as
\begin{equation}\label{cat}
\lambda^{a}_{k} \stackrel{A_{ij},e_{ijkl}} \Leftrightarrow
\psi^{a}_{k}
\end{equation}
Here we have imposed a bijective relationship among the states to
signify the feedback mechanism,though the exact relationship of
this is not clear in this case.
 Now if the memory states
$\mathcal{M}$ be envisaged as a category  with the above objects
and maps in it ( precisely speaking we have to define the
morphisms whose full understanding will require some understanding
of the nature of neuronal dynamics) then (\ref{cat}) can be
thought as a graph, which are irreflexive directed multigraphs.
Thinking the input and outputs to be a space disregarding the
internal structure, for each element $\lambda^{a}_{k}$ in the
input space we may draw an arrow from this point we get
$F(A_{ij},e_{ijkl})\lambda^{a}_{k}$ (where F is an arbitrary map in
the category) which will be essentially a point in the output
space. The important point here is that the map should preserve
the structure of the graph. There are some pertinent points to be
mentioned here, first of all in a dynamical point of view, we
have  the input space of all possible different possible states
of the system and the endomap $\delta$ of the space which takes
each state $\lambda^{a}_{k}$ in which the system will be after a
time evolution. For example if we think of an object in
$\mathcal{M}$ as a machine the input space is the set of all
possible states in which the machine can be and $\delta$ gives
for each state the state in which the machine will be if the
button is pushed once. So from these arguments the idea of
neuronal graphs seems plausible. But as we will see shortly the
graphs will have to obey the constraints of neurodynamical
equations and this may give rise to interesting conclusions as
regards the interpretation and the geometry of the functional
space. The functional connectivity of a neural system can be computed by from the covariance matrix of the graph concerned \cite{ospor}. As it can be realised the categorisation of input patterns do have a reflection in the graphs which we have constructed. There are some interesting experimental results observed  \cite{os1,os2} on macaque monkey and cat which shows that connectivity patterns have very striking overlaps with graph theoretical analysis of cortical areas.

\subsection{Neurodynamics}

It is already established that neurobiological processes may be
described by the reaction and diffusion models \cite{diffuse} In
this context it will be worthy to mention that the
Fisher-Burger's and Fokker-Plank equations are well known partial
differential equations which govern a wide variety of physical
systems. The basic model we want to describe here is based on
these prescriptions. We assume here that given an input there is a
probability that some neurons will communicate to form a graph to
yield a particular output. As it is quite well known that the
synaptic weights $A_{ij}$ do change, by the process of learning
whereby we get the evolution $$ \hat{A}_{ij} = A_{ij} + \sum
e^{ijkl} l_{ij} f(\lambda^{a}_{k},\psi^{a}_{k})$$ where $l_{ij}$ can
be labelled as learning parameters and $\beta^{k}$ to be the
strength by which the inputs are associated with the synaptic
weights by the neurons for some function. We also assume that the
inputs and outputs should be related with each other after the
synaptic weights and the neuronal strengths are integrated out.
In other words as wee will see below the models are stochastic
there are probability distributions associated with a neuronal
graph for a particular input with the corresponding output. The
main task for us is to find out the probability distribution,
which is most difficult to find out in more general
circumstances. We will try to find its nature for some simple
situations. So given this we postulate the following relation
between the output and the input states.
\begin{eqnarray}\label{inot}
R_{n}(\psi^{a}_{k}) &=& \mathcal{P}(\lambda^{a}_{k}) + \sum
q_{n}(g(\lambda^{a}_{k}))
\end{eqnarray}
 where $\mathcal{P}(\lambda, t)$ is the
Probability density function of the graph associated with a corresponding input output and the synaptic wights and
neuronal spikes, $R_{n}, q_{n}$ are
assumed to be polynomials of some general function of the output
and input states respectively in the simplest approximation. We would like to state here that in analysing the
dynamics of the graphs an appropriate spiking function of the neurons has to be incorporated, for example for a
stimulus intensity $g(\lambda)$ the spiking function could be chosen as to be the Naka-Rushton function
\begin{eqnarray}
e_{ijkl} &=& \frac{s_{max}g(\lambda)}{g_{1/2}(\lambda) + g(\lambda)};\quad g\geq 0\\ \nonumber
      &=& 0 \quad otherwise
\end{eqnarray}
Here $g_{1/2}(\lambda)$ is the stimulus intensity which produces half the maximum firing rate $s_{max}$.
  Now
the Fisher-Burger \cite{fish} scheme gives us a tool to find out a
specific differential equation for the probability distribution
function. It is quite obvious that the probability distributions
will generally be dependent on the inputs and the synaptic weights
apart from being time dependent, and we here in this scheme
assume that the synaptic weights $A_{ij}$ do depend on the inputs
and the probability distribution,the inputs in turn by a feedback
mechanism may also depend on the weights and some noise
parameters all with appropriate couplings. So the basic equations
are given as follows.
\begin{eqnarray}\label{diff}
-\frac{\partial A_{ij}}{\partial \lambda^{a}_{k}} &=&
\mathcal{P} + \beta f(\lambda^{a}_{k}) \nonumber\\
\frac{\partial A_{ij}}{\partial t} + \mathcal{P}A_{ij} &=&
-d\frac{\partial \mathcal{P}
}{\partial \lambda}\nonumber\\
\frac{\partial \lambda^{a}_{k}}{\partial t} &=&
-\frac{1}{2}\gamma e_{ijkl}f(\lambda) + \theta \nu_{i} + \phi A_{ij}
+ \zeta \psi^{a}_{k}
\end{eqnarray}
The last equation has a discrete analogue for gradient descent
algorithm in ANN's, which is given by $$\lambda^{a}_{i}(t +
\epsilon) = \lambda^{a}_{i}(t) - \frac{\epsilon}{2}\nabla{f_{i}}
+ J g(\lambda^{a}_{i})\nu_{i}(t)$$ Here it should be noted that
neuronal spikes are assumed to be nondynamical, though this may
seem to be an idealisation, it will be seen that the known models
to comply with this. Now after some straightforward algebraic
manipulations (\ref{diff}) reduces to
\begin{eqnarray}\label{prob}
\frac{\partial \mathcal{P}}{\partial t} -
\frac{\partial}{\partial \lambda}( \mathcal{P} A_{ij}) -
d\frac{\partial^{2}\mathcal{P}}{\partial \lambda^{2}} &=& -\beta
f^{'}(\lambda)\{-\frac{1}{2}\gamma e_{i}f(\lambda) + \vartheta
\nu_{i} + \phi A_{ij} + \zeta \psi^{a}_{k}\}
\end{eqnarray}
It should be noted here that (\ref{prob}) reminds us of the basic
aspects of population modelling of statistical neurodynamical
equations \cite{pop}. So essentially by some assumptions on the
neuronal variables and forming a set of coupled differential
equations between them, we are able to produce the differential
equation for the probability density function for the graphs.The
equation can be identified with the Fokker-Plank equation in the
lower order which is given by the
\begin{eqnarray}\label{fokk}
\frac{\partial}{\partial t} \mathcal{U} &=& -\eta \frac{\partial
}{\partial \lambda}a_{1}\mathcal{U} +
\frac{\eta^{2}}{2}\frac{\partial^{2}}{\partial
\lambda^{2}}a_{2}\mathcal{U}
\end{eqnarray}
Here $a_{1}, a_{2}$ drift and diffusion coefficients and
$\mathcal{U}$ is the probability distribution.With an appropriate
choice of the the coefficients and functions $f$ the equations
(\ref{prob}) and (\ref{fokk}) can be shown to be equivalent.
Though we should like to emphasize that $\mathcal{P}$ in our case
denotes the probability density associated with a graph
determined by the synaptic weights and neuronal spikes. In this
context it should be mentioned from the form of (\ref{prob}) that
as we has envisioned about the independence of weights and spikes
in the input output evolution, this may not be achieved in this
form without some furthur assumptions on those variables. Now it
will be realized that (\ref{prob}) with proper choice of
variables and parameters with some simplifying assumptions, take
the form of diffusion equation given by $$ \kappa
\frac{\partial^{2} \mathcal P}{\partial \lambda^{2}}=
\frac{\partial \mathcal P}{\partial t}$$ Now in finding the
elementary solutions of the diffusion equation we get $$ \mathcal
P = \frac{1}{4\pi\kappa t}exp{\{-(\lambda - \xi)^{2}/4\kappa
t\}}$$ Now given the form of (\ref{prob}) it is clear that the
solution will be nontrivial and this stage to make the
probability density function to be dependent on the inputs only
we require to impose constraints on the weights $A_{ij}$ and the
spikes $e_{ijkl}$. So in our case the probability density
function of the graphs the solution will take the form
\begin{equation}
\mathcal P = \mathcal C {\mathcal F}(\lambda)exp{\{-(\lambda -
\xi)^{2} a/ t\}}
\end{equation}
It should be realized that the probability density function
satisfies the information geometric constraints as has been
postulated in \cite{ama}. To give a geometric flavour in the
interpretation of the probability density function, we state the
variant of maximum-minimum principle and postulate that it may
give rise to
the information flow in the mental state space.\\
 \textbf{Theorem:}\\
 If we consider $ {\mathcal P}(\lambda,t)$ to be a continuous
 function of its arguments and is a solution of the diffussion
 equation, then $ {\mathcal P}(\lambda,t)$ attains it's extremum
 at the boundary.

Now we
would like to analyse the evolution on graphs  for the inputs by
an alternative proposal \cite{frob} by means of the
Frobenius-Perron(FP) operator and try to construct a
corresponding reachability matrix. At the first place
corresponding to each directed edge joining two lattice points we
associate a vector ${\mathcal T}$. We define the FP operator
$\hat{\mathcal O}$ on the vector by the following rule $$
\hat{\mathcal O} {\mathcal T}(e_{ijkl}, A_{ij})  = \sum_{p}
\mathcal{P} {\mathcal T}(\phi^{-g(\lambda)} (e_{ijkl}, A_{ij}))$$
where the sum is computed over all the possible paths. Taking the
Laplace's transform on the FP operator by assuming a Poisson
distribution of the paths on the functional space over the edges
of length $\textsl{l}$ and the learning parameters $l_{ij}$, the
analysis of \cite{frob1} gives rise to
\begin{eqnarray}\label{frobi}
&& \sum \int_{0}^{\infty} e^{-s
\lambda}\frac{(s\lambda)^{n}}{n!}\hat{\mathcal O}{\mathcal
T}(\phi^{-g(\lambda)} (e_{ijkl}, A_{ij})) d\lambda\nonumber\\
 &&=
\int_{0}^{\textsl{l}}e^{-s \lambda}{(s\lambda)^{n}}{\mathcal
T}(e_{ijkl} + l_{ij} log\lambda, A_{ij}))d\lambda \\\nonumber && +
\sum_{n = 1}^{\infty}\sum_{p}\int_{\textsl{l} - a\sum
e_{ijkl}}^{{\textsl{l} + a\sum e_{ijkl}}}e^{-s
\lambda}{(s\lambda)^{n}}{\mathcal T}(e_{ijkl} + l_{ij}
log\lambda, A_{ij} + {\textsl{l}}^{n}))d\lambda
\end{eqnarray}
The second integral in the rhs of (\ref{frobi}) can be written in
a tractable form by change of variables and identifying $$
{\mathcal Q}_{e e'} = U(\lambda) {\mathcal P} (s\lambda)^{n}
e^{-s\textsl l}$$ for some choice of the function $U$. Thereby it
will be seen that the sum over all the paths can be implemented
by the matrix ${\mathcal Q}^{n}$. It can be argued that
${\mathcal Q}$ may be a measure of the connection between the
edges via the sites \cite{seth} from which the covariance of the
neural system can be calculated which will give us a  clue about
the statistical information variables of the system. So
essentially the matrix ${\mathcal Q}$ encodes a great deal of
information about the neuronal systems which crucially depends on
the probability distribution of the graph.Let us see how much we
can proceed in finding out some solutions to the equation
(\ref{prob}). Let us try to construct the states in thinking that
states are formed in a stationary equilibrium in a sense that we
see the evolution of the states after long time has passed after
the input is applied. So these solutions in this case may denote
the permanent memory states of a neurological system. In finding
out the stationary states of the equation we make a proposal that
under some suitable choice of the variables, functions and initial
conditions (\ref{prob}) can be recast in the form
\begin{eqnarray}\label{nonl}
\frac{\partial}{\partial \lambda}\Big[\mathcal{A}(\lambda,
\mathcal{P}(\lambda))\frac{\partial \mathcal{P}(\lambda)}{\partial
\lambda}\Big] &=& G(\lambda, \mathcal{P}(\lambda))
\end{eqnarray}
We do assume some strict restrictions on $\mathcal{A}, G(\lambda,
\mathcal{P}(\lambda))$ for the model to bring some interesting
conclusions. Solvability of the above equation (\ref{nonl}) is
equivalent to the solvability of the generalized Hammerstein
equation \cite{hammer}
\begin{eqnarray}\label{hamm}
\mathcal{P}(\beta) + \int\mathcal{K}(\lambda, \beta ;
\mathcal{P}) G(\lambda, \mathcal{P}(\lambda)) d\lambda &=& 0
\end{eqnarray}
where $\mathcal{K}(\lambda, \beta ; \mathcal{P})$ is the kernel
of the evolution. But the essential point here after this
identification is that the solutions of this differential
equation are known to have solutions in $L^{2}[0,1]$ which may be
identified  with a reflexive Banach space with a finite Borel
measure on it \cite{ban}. In that case we may like to loosely at
this stage identify the neuronal states to be elements of a
statistical functional space which as we saw are related by the
the probability density function $\mathcal{P}$. Before discussing
the geometry of the state space, in this context it is noteworthy
to mention the saccade model in context of vector averaging can
be applied in this case or not. This case has been analyzed in
reference to investigations of visual stimuli and formation of
images off geniculate \cite{gen}. We propose that the state space
geometry can be modeled from of a neural field, $ \hat{\mathcal
J}$, whose evolution for a specific probability distribution
should be expected to give rise to the consequences. We state
below the basic equation which may give rise to the evolution of
the neural field.
\begin{eqnarray}
\dot{\hat{\mathcal J}}(\lambda, t) &=& {\mathcal P}(\lambda,
t)\ast X[\hat{\mathcal J}] + \eta(\lambda, t)
\end{eqnarray}
Here $\ast$ denotes the convolution, and $X$ is a functional of
the neural field operator.

\section{Geometry of the functional space and memory states}
Before discussing the implications of the previous analysis and
the geometry of the graph space let us turn to a simple problem
in robotics.

\begin{figure}[htbp]
\centering
    \includegraphics{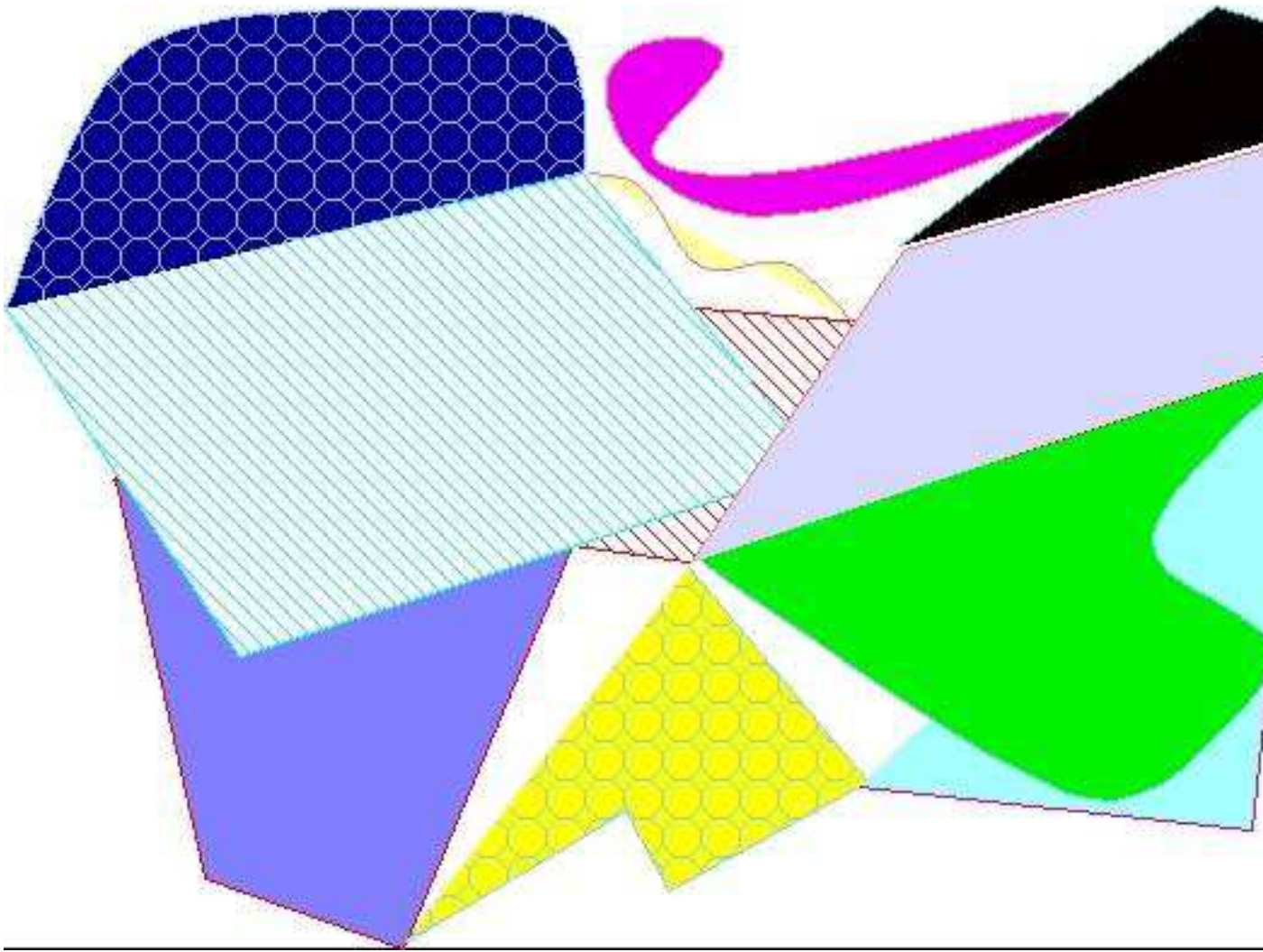}
    \caption{Cognitive States and Algebraic Surfaces}
    \label{figal}
\end{figure}

In the figure below there are 3 linked rods of lengths of 6, 4, 2
respectively with specific coordiantes. The states(positions) of the arm are determined by
the solution in $R^{6}$ to the following polynomial equations.
\vspace{0.8cm}
\begin{figure}[h]
    \centering
        \includegraphics{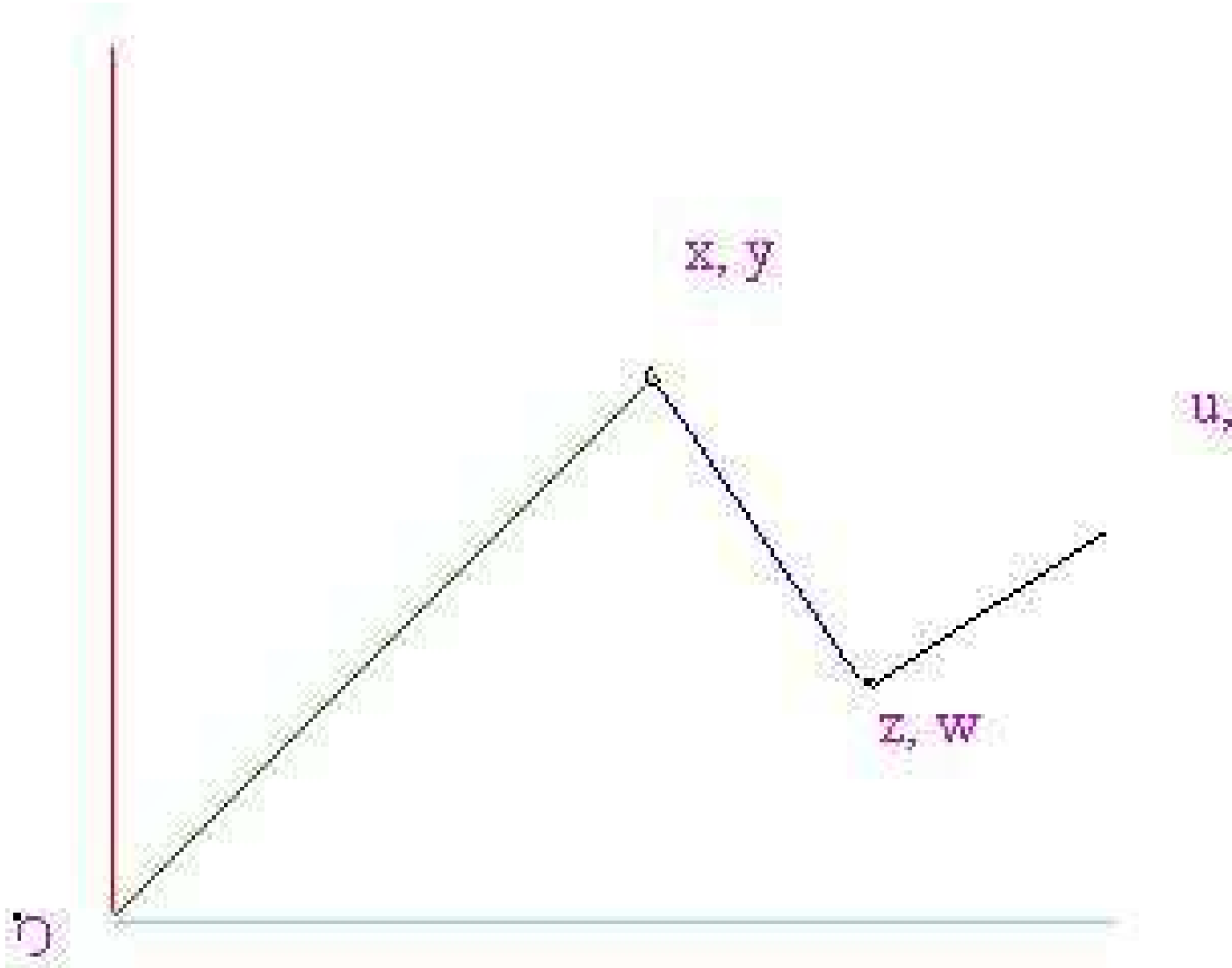}
        \caption{Geometry from Graphs }
\end{figure}
\begin{eqnarray}
x^{2} + y^{2} &=& 36 \nonumber\\
(z - x)^{2} + (y - w)^{2} &=& 16 \nonumber\\
(u - z)^{2} + (v - w)^{2} &=& 4
\end{eqnarray}
So from this simple example two things are clear, the graphs
consisting of the neurons and synapses at lattice sites is
similarly able to give rise to memory and cognitive states in a
manifold, but what is crucial to understand is that as that case
is not as simple as the previous case it is quite likely that the
structure may quite depend on the nontrivial topologies of the
input and random networks of the neuronal configurations. In a
sense different topologies of the graphs may give rise to
nontrivial manifold and geometric structure. In the diagram below
we give a simple example involving the input and output states
but with distinct topologies of the neuronal graphs which
essentially depend on the neuronal and synaptic configuration
connecting those states.
\begin{figure}[h]
    \centering
\includegraphics{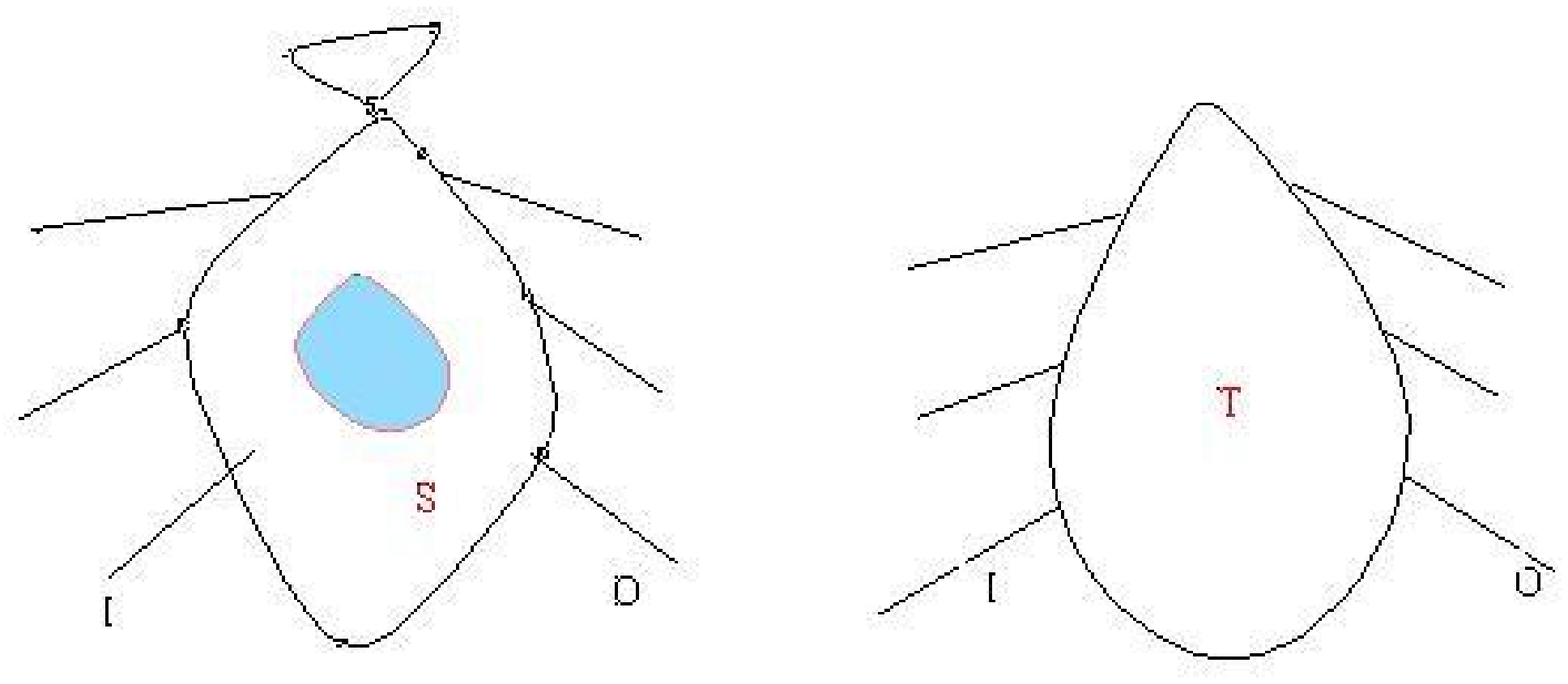}
\end{figure}
In discussing models of memory significantly there are two
aspects, short-term memory (STM), and long-term memory (LTM).
Modeling these phenomena has been taken from various possible
angles, but noteworthy among them is the importance of dynamical
systems and neural networks. What is crucial here is the choice
of  a proper network model and the memory states may be given by
the stability and attractor properties of the networks and choice
of a suitable Lyapunov function \cite{cohgross}. On the contrary
LTM is mainly concerned with storage patterns associated with
modification of synapses, through synaptic plasticity. Our
approach here is mainly composed of finding out the geometry of
the space in continuation of our preceding analysis and see how
memory can be incorporated. In this connection of gaining some
geometric intuitions in regards to brain modeling we would like
to mention an important result of Principal component analysis
\cite{pcm}. If $\lambda$ is the input stimuli, and $v$ are the
coordinates of the input vector in the state space, a map
${\mathcal K}$ is constructed such that ${\mathcal K} : \lambda
\longmapsto \textbf{\textbf{}}v$. By the process of nonnegative matrix
factorisation in which ${\mathcal K},v$ are positive definite a
real image is decomposed into the smaller units by the
decomposition of the matrix in lower dimensional eigensubspaces.
 What we propose here is that the geometry
of the surfaces of the corresponding manifold are the states of
memory, which are associated with a probability distribution. For
example the neighbourhoods of a particular point has a likelihood
of being arisen from a set of similar or simultaneous sensory
experiences, which are being correlated by the notion of a
probabilistic neighbourhood on the manifold. From the conclusions
of (\ref{hamm}) this distance may be statistical in nature. It is
quite already well known from a biological perspective that the
central nervous system evolves through natural selection of
optimal interactions with the environment. Geometrically this may
be expressed by the speculation that the neuronal circuits
matches the system of relations among objects in the external
world with a many dimensional inner geometry. Recent studies
\cite{cort}have also showed some possibility of a geometric
structure by defining distance function on the cortical areas of
the brain. But as far our studies has made us to believe that the
input and output states may be related (\ref{inot}) by the
probability distribution of the graphs which in turn is being
determined by the synaptic weights and neuronal spikes via a
differential equation which determines the structure of the
manifold in the functional space in the brain. At this point it
is quite relevant to look into the state space model which treats
the neuronal activities as a noise free dynamical system given by
the following set of equations,
\begin{eqnarray}
x(n+1) &=& \varphi(W_{i}x(n) + W_{j}u(n))\\\nonumber
 y(n) &=& {\mathcal H} x(n)
\end{eqnarray}
where $x(n)$ denote the state (q by 1 vector) of the nonlinear
system, $u(n), y(n)$ denote the input (m by 1 vector) and output
(p by 1 vector) respectively. $W_{i},W_{j},{\mathcal H}$ are the
q by q matrix, q by (m+1) matrix, p by q matrix respectively.
Here $\varphi$ is a diagonal map defined as $\varphi :
{\mathcal{R}}^{q}\rightarrow {\mathcal{R}}^{q}$
 It should be realized that the
state space model of networks may be realized in our scheme in
(\ref{inot}) by the following identification, $ \mathcal{H}
\rightarrow \mathcal{P}, W \rightarrow q, \varphi \rightarrow g$
Now it should be realized that this is an approximate and
imprecise correspondence which we state at this as a conjecture.
What we want to stress at this juncture is that we hope that our
model of geometrical picture of the state space may have the
potential of a realistic model which will obviously require
biological evidences. So if the weight factors can be integrated
out or some suitable guesses of those variables may give us a
rough idea about the manifold and the possible geometry of the
state space. In other words the output memory states are related
with a probabilistic measure in the state space in reference to
the input states, i.e. there is a distribution of the states on
the surfaces of the manifold. But as it should be realized that
inputs may form a category in terms of say systems
specificity(visual,auditory etc) or functional(colour,shape etc)
it is quite important that there would be a variety of
overlapping geometries or surfaces from a geometrical point of
view. So we would get a different probability distributions
$\mathcal{P}$ for each case. The outputs will also be arranged on
the surfaces accordingly. For example $N$ classes of signals may
occur in accordance with our model with probabilities
${\mathcal{P}}_{1}, \cdots {\mathcal{P}}_{n}$  corresponding to
distinct graph configurations. This may in turn give rise to
distinct and intersecting geometries. But how are they important
from a biological perspective? For example suppose the visual
field is exposed two sets of stimuli, a red flower and a red rocky
mountain in a succession of a large time interval. In the
language of tensor network theory these inputs may be correlated
in terms of the invariant sets of relationships, but in our
framework it is somewhat different. As we have discussed each
input gives rise to a different intersecting geometries
associated with different probability density. Now again say
after long time span suppose the subject is exposed two any one
of the stimuli. Then there is a probability that the subject may
or may not recall the other stimuli because of the intersecting
nature of the associated surfaces. In the figure below
(\ref{figal}) we have drawn various surfaces with different
colours representing multitude of cognitive states and memory.
The surfaces correspond to various stimuli distributed with a
definite probability distribution. Intersecting surfaces do show
the overlapping nature of the stimuli. The exact description and
predictability of the recall phenomena will involve a detailed
analysis of the neurodynamical equations and studies of the
corresponding geometries evolving out of these. But as we have
mentioned before whether a pattern or event will be recalled or
not is a statistical phenomena, as long as the memory states do
form a short-term memory. So as wee see interpretation or
modeling of LTM in our scheme if at all possible is not obvious
at this stage.
Some recent interesting results are worth mentioning in
connection with our investigations. Analysis of BCM theory
\cite{pop,fig} on the lateral geniculate nucleus for the visual
field by the modified hebbian learning process shows surface
formation to construct the aspects of learning and image
formation.
\begin{figure}[h]
  \centering
  \includegraphics{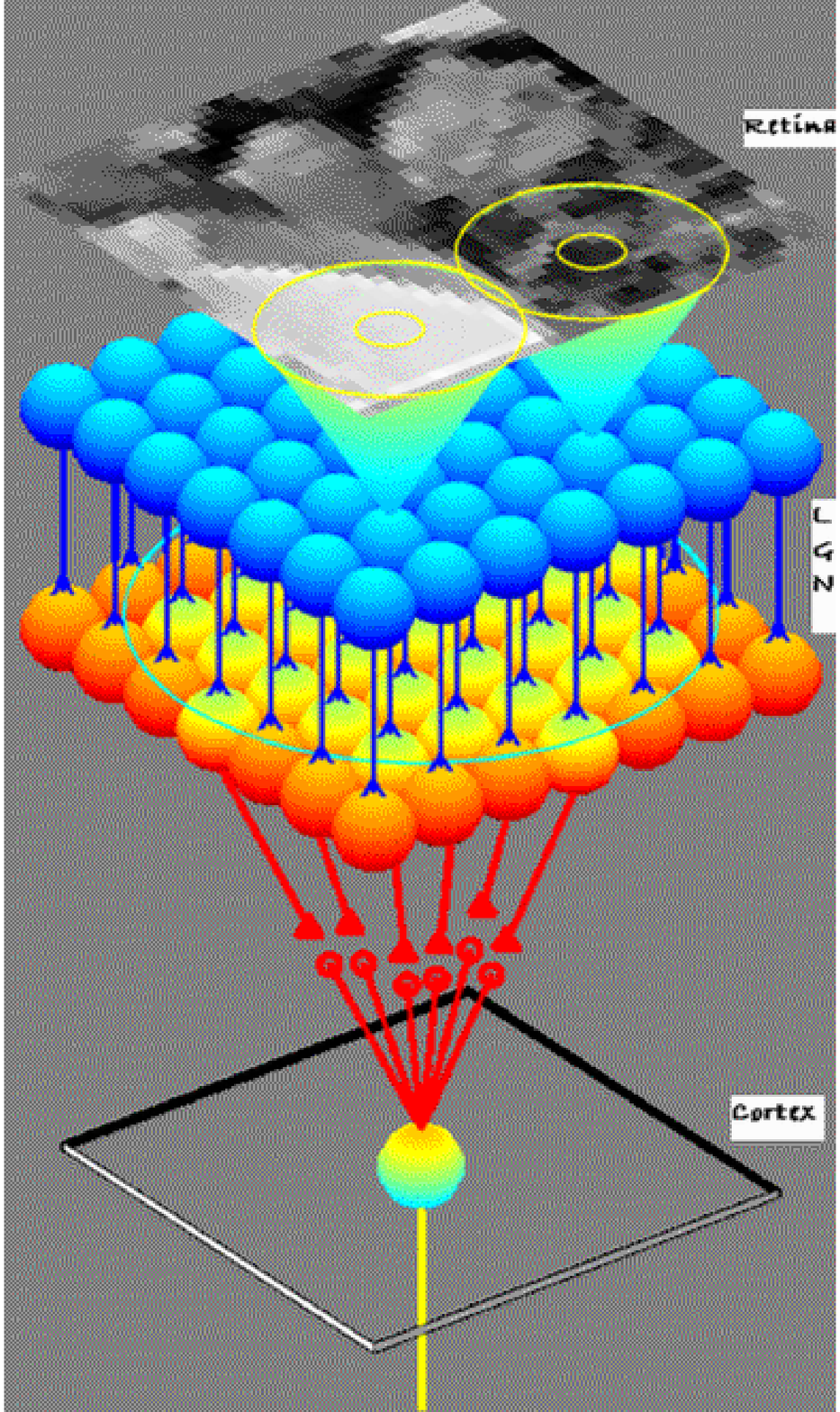}
  \caption{}\label{surf}
\end{figure}
It should be also be stressed that the self organizing maps
modeled by \cite{kohn} do exhibit topological aspects in learning
process and memory by the formation of input patterns and
synaptic weight formation. To relate with experiments we would
like to mention the recent advances in tensor maps by the
eigenvectors of the deformation gradient tensor associated with
the brain activities which transforms a time evolution of the
brain anatomy.

We sketch below some simple examples of finding out geometries of
a curve and show that what can the model proposed by us lead to
under simpler situations. In parametrizing curves we may state
that a point on the curve be given by a parametric function of
the following form
\begin{eqnarray}\label{bez}
p(u) &=& \sum_{i = 0}^{n}p_{i}f_{i}(u) \quad\quad u\in [0,1]
\end{eqnarray}
where the vectors $p_{i}$ represent the $n + 1$ vertices of a
characteristic polygon and $f_{i}(u)$ are the basis functions.
The above parametrization (\ref{bez}) is denoted as a Bezier
curve. In drawing this curve the model demands some restrictions
on the basis functions and the functions are approximated with
some polynomials. In this case a family of functions called
Bernstein polynomials satisfy those restrictions and the equation
takes the form
\begin{eqnarray}\label{bez1}
p(u) &=& \sum_{i = 0}^{n}p_{i}B_{i,n}(u) \quad\quad u\in [0,1]
\end{eqnarray}
where the polynomials satisfy
\begin{eqnarray}
B_{i,n}(u) = \frac{n!}{i! (n - i)!}u^{i}(1 - u)^{i}
\end{eqnarray}
For 3 points n = 2 and the parametrization takes the form $$ p(u)
= (1 - u)^{2}p_{0} + 2u(1 - u)p_{1} + u^{2}p_{2}$$ As it can be
understood surfaces can be parametrized along the same lines. So
the approach would be to get an idea of the geometry from the
dynamical equations and parametrize the geometry by suitable
functions for concrete predictability. In our case for example
equation (\ref{inot}) for a stationary solution of the
probability density ($\mathcal{P}$) equation (\ref{prob}) for
simple choices of the weights and neuron spikes do lead a
relationships in the input output states of the form
{\footnote{We suppress the indices here for typographical
convenience}}
\begin{equation}\label{alg}
\psi^{2} + a\psi \lambda + b\psi = \lambda^{3} + s\lambda^{2} +
t\lambda + v
\end{equation}
Here the $a,b,s,t,v$ are related to the previous coefficients.
The equation (\ref{alg}) is termed as the Weierstrass form of an
elliptic curve \cite{ell}. So we may get a specific geometry
${\mathcal M}_{i}$ and the memory states for each stimulus are
embedded in this geometry with a probability density function
${\mathcal P}_{i}$.

\section{Concluding Remarks}
The above analysis clearly indicates that neuronal circuits may
give rise to a neuronal geometry by the realization of graphs on
a function space. The existence of a distance function and a
probabilistic metric tensor unlike some previous analysis
\cite{cort} is not clear from the discussions presented here. But
it is somewhat clear that the geometry is of probabilistic
nature. The nature of the paper has been developing some new
ideas and proposing its realizations. We hope that this approach
may give rise to some interesting conclusions in deeper
theoretical and experimental analysis as regards information
processing, and theories of cognition, and memory.

\section*{Acknowledgements}
The author wishes to thank S.Roy (PAMU),ISI for extremely helpful
and stimulating discussions in aspects of neuromanifold
theory.The author also extends his sincere gratitude to SINP
hospitality for being able to complete this work.


\begin{thebibliography}{99}
\bibitem{amari}S.I.Amari and H.Nagaoka (2000) {\it Amer.Phys.Soc}
\bibitem{hop}J.Hopfield (1982) {\it Proc.Natl.Ac.Sci} 79 2554
\bibitem{amit}D.J.Amit, H.Gutfreund and H.Sompolinsky (1985)
{\it Phys.Rev.Lett} 55 1530
\bibitem{sch}E.Schwartz, R.Desimone, T.Albright and C.Gross (1983)
{\it Proc.Natl.Ac.Sci} 80 5776
\bibitem{rock}E.T.Rolls and A.Treves (1998) ``Neural Networks and
Brain function'' (OUP)
\bibitem{hof}D.Hubel and T.Wiesel (1962) {\it Scientific American}
241 130
\bibitem{gross}S.Grossberg (1978) {\it Prog.Th.Biology} Vol 5 233.
\bibitem{bak}A.M.P Andrew (1969) { \it Prog.Cybn} Vol1 359
\bibitem{top}E.Alfinito and V.Itello (2000) quant-ph/0006066
\bibitem{top1}G.Recanzone, M.Merzenich, W.Jenkins, A.Kamil and
H.R.Dinse (1992) {\it Jour. Neurophys} 67 1031
\bibitem{band}P.A.Bandettini, A.Jesmanowicz, E.C.Wong and J.S.Hyde
(1993) {\it Mag.Res.Med} 30 161
\bibitem{link}A.Longtin, A.Bulsara and F.Moss (1991) {\it
Phys.Rev.Lett} 67 656
\bibitem{hamm}S.Hagan, S.Hameroff and J.A.Tuszynski (2002) {\it
Phys.Rev E} 65 61901
\bibitem{ras}M.Riani and E.Simonotto (1994) {\it Phys.Rev.Lett} 72
19
\bibitem{eeg}P.Mansfield, R.Coxon and P.Glover (1994) {\it
J.Comp.asst.Tomogr} 18 339
\bibitem{crit}D.J.Amit (1989) ``Modelling Brain function'' CUP
\bibitem{pib}K.H.Pribram (1991) ``Brain and Perception'' LEA
\bibitem{fre1}W.J.Freeman (1991) {\it Scient. Amer} 264 78
\bibitem{ur}G.Bernroider (2003) {\it NeuroQuant} 2 163
\bibitem{ya}Y.Zhou, J.H.Morais-Cabral, A.Kaufman and R.Mackinnon
(2001) {\it Nature} 335 311
\bibitem{pen}S.Hameroff and R.Penrose (1996) {\it J.Cons.Studies}
3 36
\bibitem{pv}H.Haken (1978) ``NonLinear Phase Transitions in
Physics,Chemistry,Biology'' Springer-Verlag
\bibitem{qc}S.Panzeri and A.Treves (1996) {\it Comp.Neur.Sys} 7 87
\bibitem{roy}S.Roy and M.Kafatos (2004) {\it Forma} 19 69
\bibitem{mem}E.Miller and R.Desimone (1994) {\it Science} 254 1377
\bibitem{map}T.Kohonen (1982) {\it Biol.Cybntcs} 43 59
\bibitem{cog}J.Hertz, A.Krogh and R.G.palmer (1991) ``Introduction
to the theory of neural computation'' Addison Wesley
\bibitem{comp}A.von.Ooyen (2001) {\it Compu.Neur.Sys} 12 R1
\bibitem{comp1}S.Grossberg (1978) {\it J.Theo.Biol} 73 101
\bibitem{hebb}D.Hebb (1949) ``The Organisation of behaviour''
Wiley
\bibitem{markov}J.Collins, C.Chow and T.Imhoff (1995) {\it
Phy.Rev E} 52(4) R3321
\bibitem{roy1}G.Bernroider and S.Roy (2004) {\it Forma} 19 55
\bibitem{catg}R.O.Duda and P.E.Hart (1973) ``Pattern
classification and scene analysis'' Wiley
\bibitem{mon}R.Montague, J.Gally and G.M.Edelman (1991) {\it
Cer.Cortex}  1 199
\bibitem{hodge}D.J.Watts and S.H.Strogatz {\it Nature} (1998) 393 440
\bibitem{meg}P.L.Nunez (1981) ``Electric fields of the brain: the
neurophysics of EEG OUP
\bibitem{wav}V.Y.Vasilev (2001) {\it Proc.Natl.Ac.Sci} 109 2016
\bibitem{ospor}O.Sporns, G.Tononi and G.Edelman (2000) {\it
Neural Netw} 13 909
\bibitem{os1}G.Tononi, A.R.Mcintosh, D.P.Russel, and G.M.Edelman (1998){\it
Neuroimage} 7 133
\bibitem{os2}J.W.Scannel and M.P.Young (1993) {\it Curr.Biol} 13
1991
\bibitem{diffuse}W.Horsthemke (1999) {\it Phys.Lett A} 263 285
\bibitem{fish}J.Fort and V.Mendez (2002) {\it Reprts.Prog.Phys} 65
895
\bibitem{pop}E.L.Bienenstock, L.N.Cooper and P.W.Munro (1982) {\it
J. Neurosc} 2 32
\bibitem{ama}S.Amari (1998) {\it Neur.Comp} 10 251
\bibitem{frob}F.Barra and P.Gaspard (2000) {\it J.Stat.Phys} 101
\bibitem{frob1}P.Gaspard and X.J.Wang (1993) {\it Phys. Reports}
235 321
\bibitem{seth}A.K.Seth and G.M.Edelman (2004) {\it Theoretical
Neuroanatomy} Springer
\bibitem{hammer}S.Itoh (1978) {\it Math.Ann}  236 133
\bibitem{ban}M.A.Krasnoselskii(1964) ``Topological methods in the
theory of nonlinear differential equations'' Pergamon
\bibitem{gen}V.Braitenberg, A.Scutz (1998) {\it
Cortex,Connectivity and geometry of neuronal activity} Springer
\bibitem{cohgross}W.A.Little and G.L.Shaw (1978) {\it Math.Biosc}
39 281
\bibitem{pcm}R.Kotter (2001) {\it Phil.Trans.Roy.Soc.Lond} B 356
1111
\bibitem{cort}A.Pellionisz and R.llinas (1982) {\it Proc.Jap}  394
\bibitem{fig}R.M.Gray (1984)
{\it IEEE} 1 4
\bibitem{kohn}T.Kohonen (1996) {\it Biol.Cybern} 75 281
\bibitem{ell}W.Fulton (1984) ``Intersection Theory''  Springer
-Verlag




\end{thebibliography}
\end{document}